\documentclass[Afour,SageV,times]{sagej}

\usepackage{moreverb,url}
\usepackage[colorlinks,bookmarksopen,bookmarksnumbered,citecolor=red,urlcolor=red]{hyperref}

\newcommand\BibTeX{{\rmfamily B\kern-.05em \textsc{i\kern-.025em b}\kern-.08em
T\kern-.1667em\lower.7ex\hbox{E}\kern-.125emX}}

\newcommand{\D}[1]{\textcolor{red}{}} 

\def\volumeyear{2024}

\begin{document}

\runninghead{Achitouv, Chavalarias and Gaume}

\title{Testing network clustering algorithms with Natural Language Processing}

\author{Ixandra Achitouv\affilnum{1}, David Chavalarias\affilnum{1,2} and Bruno Gaume\affilnum{1}}

\affiliation{\affilnum{1} ISC-PIF - Institut des Systèmes Complexes - Paris Ile-de-France, CNRS\\
\affilnum{2}CAMS - Centre d'Analyse et de Mathématique sociales }

\corrauth{ISC-PIF, 113 Rue Nationale, 75013 Paris France}

\email{ixandra.achitouv@cnrs.fr}

\begin{abstract}
The advent of online social networks has led to the development of an abundant literature on the study of online social groups and their relationship to individuals' personalities as revealed by their textual productions. Social structures are inferred from a wide range of social interactions. Those interactions form complex -- sometimes multi-layered -- networks, on which community detection algorithms are applied to extract higher order structures. The choice of the community detection algorithm is however hardily questioned in relation with the cultural production of the individual they classify. In this work, we assume the entangled nature of social networks and their cultural production to propose a definition of \textit{cultural based} online social groups as sets of individuals whose online production can be categorized as social group-related. We take advantage of this apparently self-referential description of online social groups with a hybrid methodology that combines a community detection algorithm and a natural language processing classification algorithm. A key result of this analysis is the possibility to score community detection algorithms using their agreement with the natural language processing classification. A second result is that we can assign the opinion of a random user at $>85\%$ accuracy.

\end{abstract}

\keywords{community detection, natural language processing, social network, classification}

\maketitle

\section{Introduction}
The advent of online social networks has led to the development of an abundant literature on the study of online social groups and their relationship to individuals' personalities as revealed by their textual productions \cite{golbeck2011predicting}\cite{gosling2011manifestations}. Social structures are usually represented as graphs where the individuals are the nodes and the links can represent a wide range of social interactions.

These graphs feature community structures (CS), which can be defined as subsets of nodes within which interactions are denser than with the rest of the network. These communities, identified through community detection algorithms,  provide insight into the social structures of the online social network under study.

Identifying relevant community structures within complex networks is however a challenging task and most algorithms are based on the intrinsic optimization of scoring functions that are often not comparable. Indeed, given a graph with a set of nodes and edges, enumerating all possible communities is an NP-Complete problem \cite{GAREY1976237}. Hence many Community Detection Algorithms (CDA) aim to optimize a quality function or scoring function that is not universal, and do not necessarily find the optimal communities (when they are known) \cite{chakraborty2016metrics, gaume2024unified}. 

A wide variety of CDA have been proposed based on different scoring functions \cite{yang2012defining}. These functions can maximize internal connectivity between nodes (e.g. density, number of edges between members of a community), external connectivity (e.g. number of edges per nodes that point outside a cluster \cite{Radicchi_2004}), or both (e.g. the fraction of total edge volume that points outside the cluster \cite{Jianbo}). Alternatively, the modularity is another scoring function introduced in \cite{NewmanGirvan} and is defined as the difference between the number of edges between nodes in a community and the expected number of such edges in a random graph with the same degree sequence. Assessing the quality of these CDA (validation procedure) is usually performed using `ground truth' known communities or by generating artificial graphs \cite{Lancichinetti_2008}.

However one could wonder if we can introduce a scoring function independent of the properties of the network, which would rely on the actual exchanges of information among the nodes such as the content of messages between users that are identified as nodes. This is the scope of this article -- using a Natural Language Processing Classification Algorithm (NLPCA), we introduce a new scoring function for comparing the outputs of CDA on online social network data. In addition one might wonder how this scoring maps into understandable social communities, and how can a random user within such communities be precisely categorized. 

We will address these questions and introduce a scoring function based on a trade-off between precise categorization of users and coverage of the user categories (recall), among a reduced number of communities. These textual classifications/CDA have  direct applications in social science where we want to analyse meaningful communities. A second key result of our analysis is the high accuracy in classifying a random user opinion based on NLPCA. Our approach is different than other works \cite{frank2018inferring,van2018gender,schwartz2013personality, ferrara2014user,le2019detecting} that use NLPCA without fine tuning models on the CDA. In our case, we use a CDA on a given interaction network to define social groups whose cultural production is used to train a NLPCA. Then we use the performance of the NLPCA in the categorization of new individuals based on their online cultural production, to assess the quality of the CDA. This makes it possible to assess in an integrated way methods to define \textit{cultural based} online social groups \textit{i.e.} social groups whose online interactions are consistent with the cultural production of their members. Coherent methods are those for which the choice of the interaction networks to analyse, and the associated community detection algorithm, lead to the highest prediction scores in terms of user categorization by the best possible NLPCA.

As a case study, we perform this assessment on Twitter data related to climate change to identify the different types of social groups that debate on this topic. We show that we can identify pairs of CDA/NLPCA that predict the opinion of a random user at $>85\%$ accuracy with $\le 3$ succinct sentences. We hence provide a pipeline to identify the best community structure algorithms and identify their optimal parameters (if any).

This work is organised as follows: first, we introduce the social network considered for this study and the CS algorithms cases selected to illustrate our analysis, as well as some fundamental concepts of natural language processing. We then describe our methodology and present our results, and finally we conclude our analysis. 

\section{Communities in the Twitter Social Network }\label{sec_studycase}

\subsection{Climate change related tweets}
\begin{figure}
    \centering
    \includegraphics[width=0.99\linewidth]{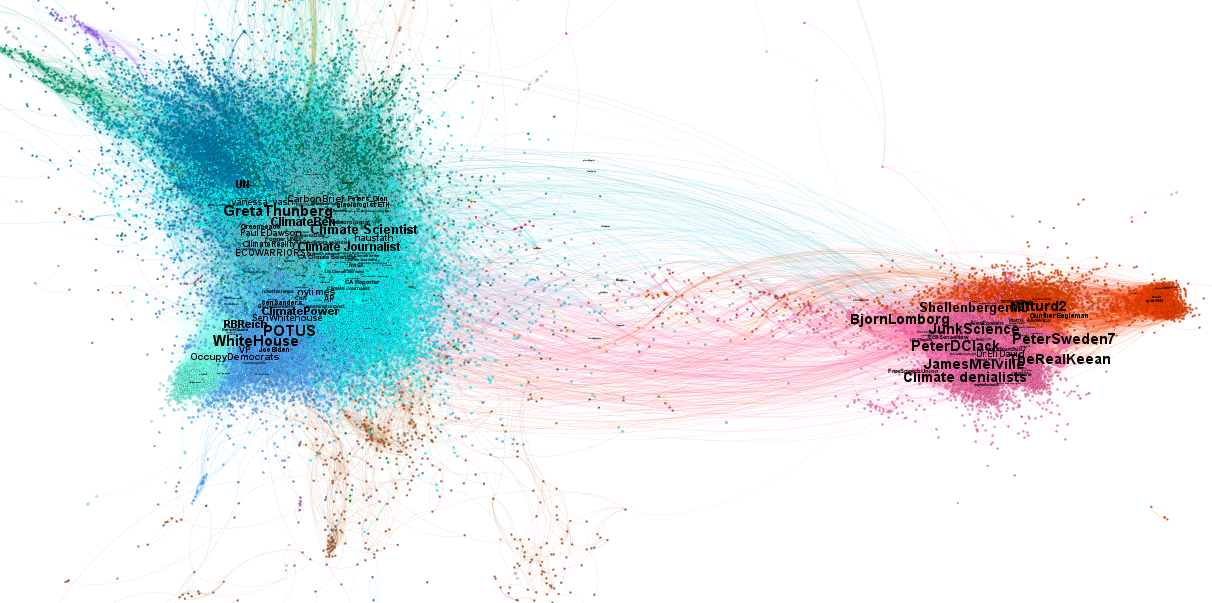}
    \caption{Visualisation of climate change related tweets from 2022-07-01 until 2022-10-30, where colors represent different communities: cold/warm colors correspond to pro-climate/denialist users respectively. In total there are 29347 accounts (nodes) and 361559 retweets (edges) among those accounts.}
    \label{fig:enter-label}
\end{figure}

We used the data from the Climatoscope project \cite{chavalarias_new_2023} to extract the retweet network of online Twitter (now `X') discussions about climate change over the year 2022. The Climatoscope project used Twitter's track API, which allowed to capture all tweets mentioning a given expression, collecting tweets based on a list of several dozens of English and French keywords related to climate change. This data collection was not exhaustive but represents a sufficiently large and diverse sample of climate change Twitter debates to understand the diversity of the social groups involved in them. Over the year 2022, 57M tweets have been collected, 32.1M of them being retweets.

We computed the retweet network, from 2022-07-01 until 2022-10-30, where the weight of an edge between two accounts equals the maximum number of retweets in either direction. The resulting network, weighted and undirected such that it can be processed by most CDA, was made of roughly 226,000 nodes and 430,000 edges. To identify the English speaking communities, we ran a standard Louvain community detection \cite{Blondel_2008} on this graph. We removed loosely connected nodes with degree strictly lower than 3, and kept the largest English-speaking communities only, pro-climate and denialist. 

The resulting graph was made of 30,000 nodes and 362,000 links. On one hand, these included international organizations (UN, COPX, UNICEF, NASA, etc.), climate activists (Greta Thunberg, Greenpeace, etc.) and communities centered on US Democrats: the left wing of the Democratic Party - around Bernie Sanders and Alexandria Ocasio-Cortez - and the {mainstream} Democratic party around Joe Biden, Kamala Harris and Barack Obama. On the other hand, the denialist communities feature Donald Trump's supporters and `Make America Great Again' (MAGA) Republicans, accompanied by other right-wing political leaders such as those of the United Kingdom Independence Party, and communities of influencer `experts' in climate science, who have their own audience and are densely connected to each other. It is in this latter denialist community that one finds accounts like JunkScience (Steve Milloy) notoriously supported by the fossil fuel industries such as the Heart Land Institute or the Competitive Enterprise Institute.

Our network is publicly available at \url{https://github.com/IxandraAchitouv/CDA_NLPCA.git}, with user IDs anonymized.

\subsection{Community Detection Algorithms (CDA)}
In what follows we explain how NLPCA can be used as a `ground-truth' community structure to test CDA, and illustrate our results with three known CDA. 

\subsubsection{The Louvain algorithm. \label{secLouv}}
The Louvain CDA is a method to extract non-overlapping communities from large networks \cite{Blondel_2008}. It runs in time $O\;[ n\cdot \log n]$ where n is the number of nodes in the network. In the Louvain method, small communities are found by optimizing modularity locally on all nodes, then each small community is grouped into one node and the first step is repeated, where the modularity is the difference between the number of edges between nodes in a community and the expected number of such edges in a random graph with the same degree sequence \cite{NewmanGirvan}. It is defined as a value in the range  $\displaystyle [-1/2,1]$, 
\begin{equation}\label{modularity}
 Q = \sum_{i=1}^{m}\frac{w_{ii}}{w} - \frac{w_{i}^{\mathrm{in}}w_{i}^{\mathrm{out}}}{w^2}.
\end{equation}
Here $w_{ii}$ is the total weight of links starting and ending in module $i$, $w_{i}^{\mathrm{in}}$ and $w_{i}^{\mathrm{out}}$  the total in- and out-weight of links in module $i$, and $w$ the total weight of all links in the network. To estimate the community structure in a network, Eq.~\ref{modularity} is maximized over all possible assignments of nodes into any number $m$ of modules. In \cite{Lambiotte_2014}, a stability criterion of a network partition is introduced, a measure of its quality defined in terms of the statistical properties of a dynamical process taking place on the graph. The time-scale of the process acts as an intrinsic parameter that uncovers community structures at different resolutions. This method has been applied to find multi-scale partitions in the Louvain algorithm with a scale that we refer as `c' in what follows. 

\subsubsection{BEC.\label{BBECalgo}}

\cite{gaume2024unified} propose a clustering method based on the optimization of the precision and recall (F-score) of a clustering relative to its ability to classify the edges of a network into clusters. It runs as an agglomeration process that reviews each edge of a network only once and merges the clusters of their nodes if this operation does not decrease the F-score. Hence there is a natural scale that is introduced, $s$ which corresponds to the trade-off between precision and recall. It runs in time $\sim O\;[ 3\vert E\vert ]$ where $\vert E\vert$ is the number of edges in the network. 

\subsubsection{Infomap.}
Infomap reveals community structure in weighted and directed networks. The method decomposes a network into modules by optimally compressing a description of information flows on the network \cite{Rosvall_2008}. It is a two-level description that allows to describe the path of a random walk visiting nodes, using fewer bits than a one-level description. Basically when a walk is within a module (cluster of nodes), it spends long periods of time there. To optimize the compression, Infomap uses the map equation $L(\mathsf{M})$ which gives the average number of bits per step that it takes to describe an infinite random walk on a network partitioned according to  $\mathsf{M}$:
\begin{equation}\label{map}
L(\mathsf{M}) = q_{\curvearrowright} H(\mathcal{Q}) + \sum_{i=1}^{m}p_{\circlearrowright}^iH(\mathcal{P}^i).
\end{equation}

where M is a module partition among $m$ modules. The first term corresponds to the entropy of the movements between modules and the second is the entropy of movements within modules. Each entropy is weighted, with $q_{\curvearrowright}$ being the probability that the random walk switches modules on any given step and $p_{\circlearrowright}^i$ the fraction of intra-module movements occurring in module $i$, plus the probability of exiting module $i$ such that $\sum_{i=1}^mp_{\circlearrowright}^i=1+q_{\curvearrowright}$.

The running time of Infomap depends on several factors, including the size of the network (number of nodes and edges) and the structure of the network (such as the density and distribution of edges).

\subsection{Assigning Tweets to Categories\label{secNLP}}

\subsubsection{Natural Language Processing (NLP).}
Natural Language Processing is one of the key pillars of artificial intelligence that enable to understand, interpret, and generate human language in an automated way. NLP algorithms are designed to understand and interpret the meaning of text data by mapping text into high dimensional mathematical vectors (this mapping is refereed as the embedding). If two vectors are close to one another in this space, it means that the two words or sentences are closely related. In the state of the art of NLP, this mapping relies on machine learning algorithms (see \cite{achitouv2023natural} for a summary of the NLP techniques over the last decade), including transformer models~\cite{Vaswani}, to better learn latent semantic links between words in a sentence. Bidirectional Encoder Representations from Transformers (BERT), is a natural language processing method based on the transformer architecture \cite{BERT}. It represents a significant advancement in the field of language understanding and has been widely adopted for various language-related tasks as it is designed to capture contextual information from both the left and right context of words in a sentence. BERT can be fine-tuned for specific tasks, such as sentiment analysis, question answering, or named entity recognition. This fine-tuning process adapts the model to more specialized tasks and datasets. For this analysis we use the freely available BERTweet model \cite{nguyen2020bertweet} which is a fine-tuned model of BERT trained using a large corpus of tweets, allowing it to analyse and generate representations for Twitter-specific language elements, such as hashtags, mentions, and emoticons. 

\subsubsection{The classification algorithms.}

In order to classify the tweets into categories (selected communities), we test several algorithms and select the best performing ones on our test datasets. Those are: (a)-linear classifiers with Stochastic Gradient Descent (SGD) learning \cite{bottou2010large}; (b)-Support Vector Classification (SVC) \cite{cortes1995support}; (c)-Multi-layer Perceptron Classifier (MLPC) \cite{rumelhart1986learning}, with ReLU-activation and hidden layer sizes of (5,2), and (d)- a Random Forest classifier (RFC) \cite{breiman2001random}.

\subsubsection{The weighted ensemble model.}
In order to reduce the variance of the errors we use all 4 classifiers with a weight according to their performance. For a given tweet, we weight the output of classifier (a) and (c) once ($w_a=1$, $w_c=1$), twice the output of classifier (d) ($w_d=2$) and 3 times more the output of classifier (c) ($w_c=3$). Then we assign the tweet to the category that has the largest number count among the 7 dimensional vector prediction ($\sum_i w_i=7$).

\section{Methodology\label{sec_metho}}
In what follows every node is a user and every weighted edge corresponds to the number of tweets user $i$ has re-tweeted from user $j$. 
For every CDA under review, we convert the directed network to an undirected graph and proceed as follows: 

\subsection{Step 1: attribution of a CDA categories.}
 We run the CDA on the network resulting in N community structures that we map to $N_{cut}<N$. To do so we keep the first $N_{cut}-1$ communities with the largest number of users and we assign all other users to community $i=N_{cut}$. This first step is essential for two reasons. First, the classification algorithms require a minimum size for the training set to perform accurate classification. When there are not enough user in a community, the number of tweets is too low. Second, depending on the parameters of the CDA, we generally have a number of communities that varies from a few hundreds to a few thousands (the limit being the number of nodes). These large numbers are not what a human interpretation of the community can handle if one is interested in classifying opinions of users. For instance in the climate tweets some communities can be climate denialists, pro-climate activists, pro-climate scientists, denialists advocating for fuel energy, etc. Larger numbers of communities with only a couple of users each are not particularly interesting to understand group dynamics. Hence the last category $i=Ncut$ is a ``catch-all term" one and is not used later on to evaluate the accuracy of the predictions.

\subsection{Step 2: splitting users into training, testing lists.}
For each user we have its category attribution from the CDA. In order to run a ML classification on tweets, we split users into training and testing sets. For the training data we consider a fraction of users that corresponds to the most influential people. All the others are assigned to the testing set. This is motivated by (a) the fact that we don't need CDA to identify the most famous users (anchors) in a social network and (b) anchor tweets are often retweeted by many users, hence performing a ML classification on these tweets can help us find the community a random user belongs to. To select the most influential users, we could use a few metrics (page-rank, eigenvector centrality, degree centrality, etc) or select by hand users that we believe have representative ideas. In what follows we consider influential people as users with an eigencentrality greater than $.75$-quantile, which is a measure of the influence of a node in a connected network \cite{shaw1954group, bottou2010large}.

\begin{figure}
    \centering
    \includegraphics[width=0.8\linewidth]{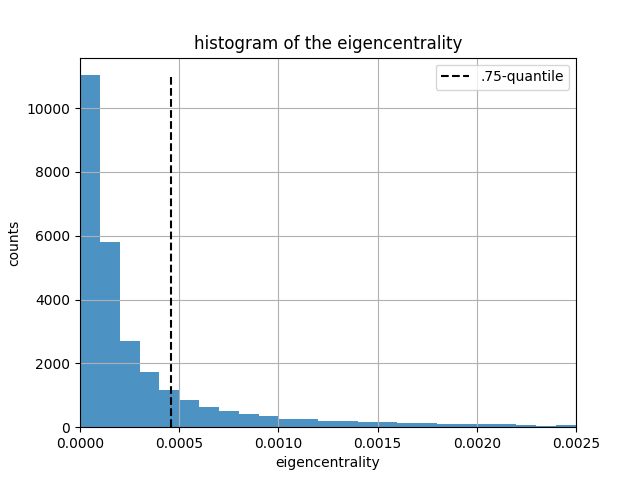}
    \caption{Histogram of user eigencentrality in our network. The vertical line corresponds to the $.75$-quantile, which corresponds to the cut between anchors and tested users. }
    \label{fig:pdfeng}
\end{figure}

Fig.\ref{fig:pdfeng} displays the histogram of user eigencentrality. In social networks, power-law distributions are often associated with degree distributions, where a few nodes (users) have significantly more connections than others. However, eigenvector centrality takes into account not just the number of connections but also the importance of those connections. The vertical line corresponds to the $.75$-quantile. Users on the right-hand side are selected as anchors for training and users on the left-hand side are used to perform the testing. 

We end up with 7,330 users (out of 29,000) for the anchors (training set) and the other users are assigned to the testing set. At the end we obtain 1,467,399 tweets from the anchors (for the training set) and 1,948,232 tweets from the other users (for the testing set).\footnote{We note that the total number of anchors does not impact our analysis when this number is divided by at least 3 times. The only issue in reducing the number of anchors is that we obtained sometimes a smaller amount of tweets than our threshold for training a given category of a given CDA. This is particularly true when the number of CS is greater than a few thousand.} Then we select every tweet of all anchors flagged into $i\in[1,N_{cut}]$ categories in our training sets such that every tweet has a category associated to it, given by the category of the anchor who emitted it (identified in Step 1 above). 

\subsection{Step 3: performing a NLPCA classification training.\label{labstep3}}
In order to have unbiased training and testing datasets, given a CDA we select a fixed number of tweets per category$_i$: $N_{train}^{i}$ for $i \in [1,N_{cut}]$. For instance, we find that for $N_{cut}=5$, we have obtained a convergence of the accuracy of the classification for $N_{train}^{i}=25,000$ tweets for the training sample of category $i$. We also select the same number of testing tweets for each category. Then we run the NLP classification algorithms described previously.

\subsection{Step 4: evaluation of CDA performance.}
The final step is to evaluate each CDA classification of a user based on its agreement with the NLPCA. For each tested tweet we have both the category of the CDA and the category of the NLPCA. Each tweet is associated with a user, so we can reconstruct the NLPCA classification of the user using the k-tweets this user made in the testing set, with $k \in [1,N]$ and $N$ is an integer. His category $i$ corresponds to the maximum count of his tweet flagged as $i$ by NLPCA.

\section{Global Results \label{sec_resu}}

\subsection{Can we precisely classify a random user?}
\begin{figure}
    \centering
    \includegraphics[width=1.\linewidth]{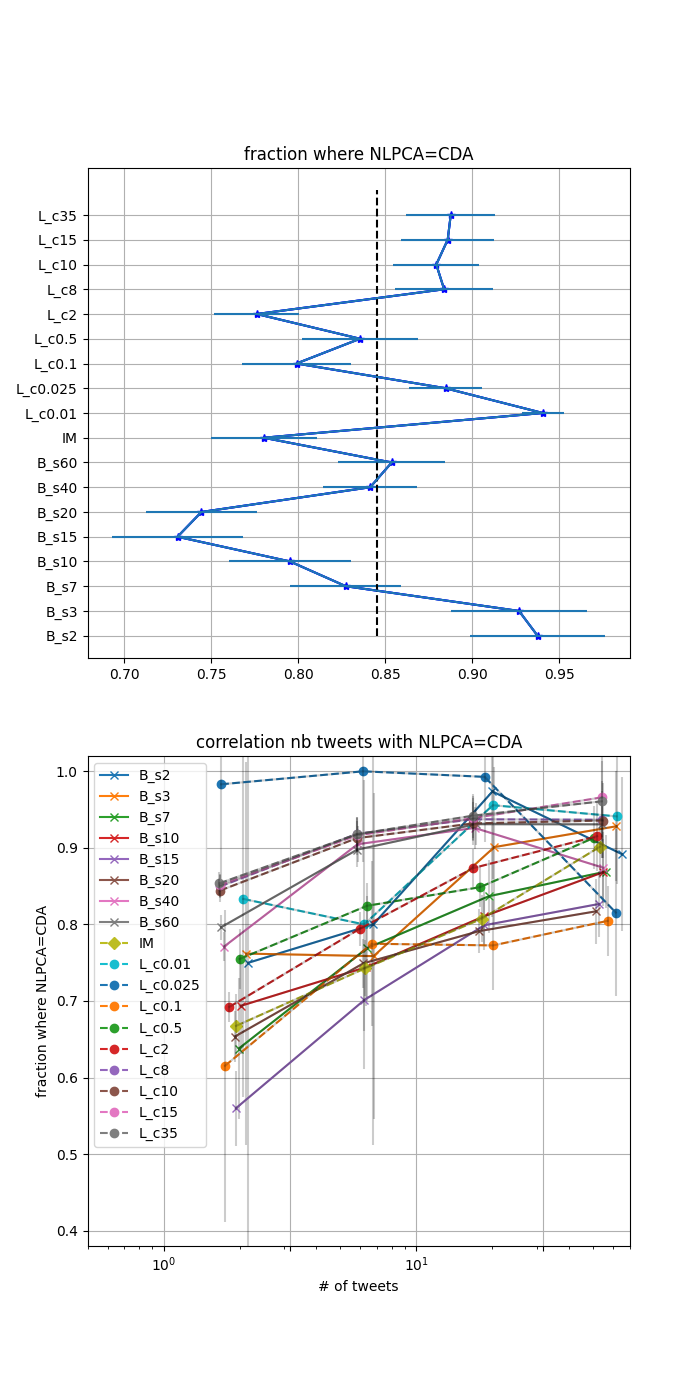}
    \caption{CDA accuracy based on its agreement with the NLPCA. $L_c$ corresponds to Louvain with parameter c, $B_s$ corresponds to BEC with parameter s and IM to Informap. Top panel:  fraction of users where the NLPCA agrees with CDA regardless of the number of tweets. The error bars are 1-sigma deviation computed by Jackknife resampling. The vertical dotted line corresponds to the mean of the accuracy for all CDA we consider. Lower panel: fraction of user where the NLPCA agree with CDA as function of the number of tweet a user made in the testing set. }
    \label{fig:Fig1}
\end{figure}

If one is interested in classifying a random user, the precision of the CDA needs to be privileged. In Fig~\ref{fig:Fig1} (top panel) we display the accuracy of the CDA based solely on the agreement with the NLPCA. Using the testing dataset with $25,000 \times N_{cut}-1= 100,000$ tweets we show on the left panel, for all CDA, the fraction of users that agrees with the NLPCA. The error bar corresponds to 1-sigma statistical deviation computed from a Jackknife resampling while the vertical dotted line corresponds to the average agreement using all CDA, here it is $85\%$ agreement.

From this figure, we deduce that the best re-scaled modularity parameters for the Louvain correspond to $c< 0.025$ while the optimal parameter for the BEC is $b< 7$. For these parameters the CDA classifications agree with the NLPCA at a precision $>90\%$, remarkably. 

On the lower panel of Fig~\ref{fig:Fig1}, we display the fraction of user where the NLPCA agrees with the CDA as a function of the number of tweets made by users in the testing set. The binning of tweets number is logarithmic. The first bin corresponds to $[1,3]$ tweets, second to a number between $[4,10]$, third to $[11,31]$, and then $\ge 32$ tweets. As expected, it is more challenging to classify a user based on a few tweets compared to a larger number of tweets. However it is still quite impressive to see an agreement at $\sim 85\%$ for several CDA, considering that $[1,3]$ tweets can characterize an unknown random user from the training set. Then we observe that the agreement increases with the number of tweets. The decrease of some curves is not statistically significant as the number of users who posted more than 15 tweets reduces to less than 10 in some cases. Poisson errors are displayed in light grey on the figure.

\subsection{Precision vs. coverage: can we categorize most users?}
A key point to address in the CDA is the number of users covered by our selected categories. Indeed, when the percentage of users in our selected categories is low, one might doubt the utility of  communities towards understanding the opinion of most users. 
\begin{figure}
    \centering
    \includegraphics[width=1.\linewidth]{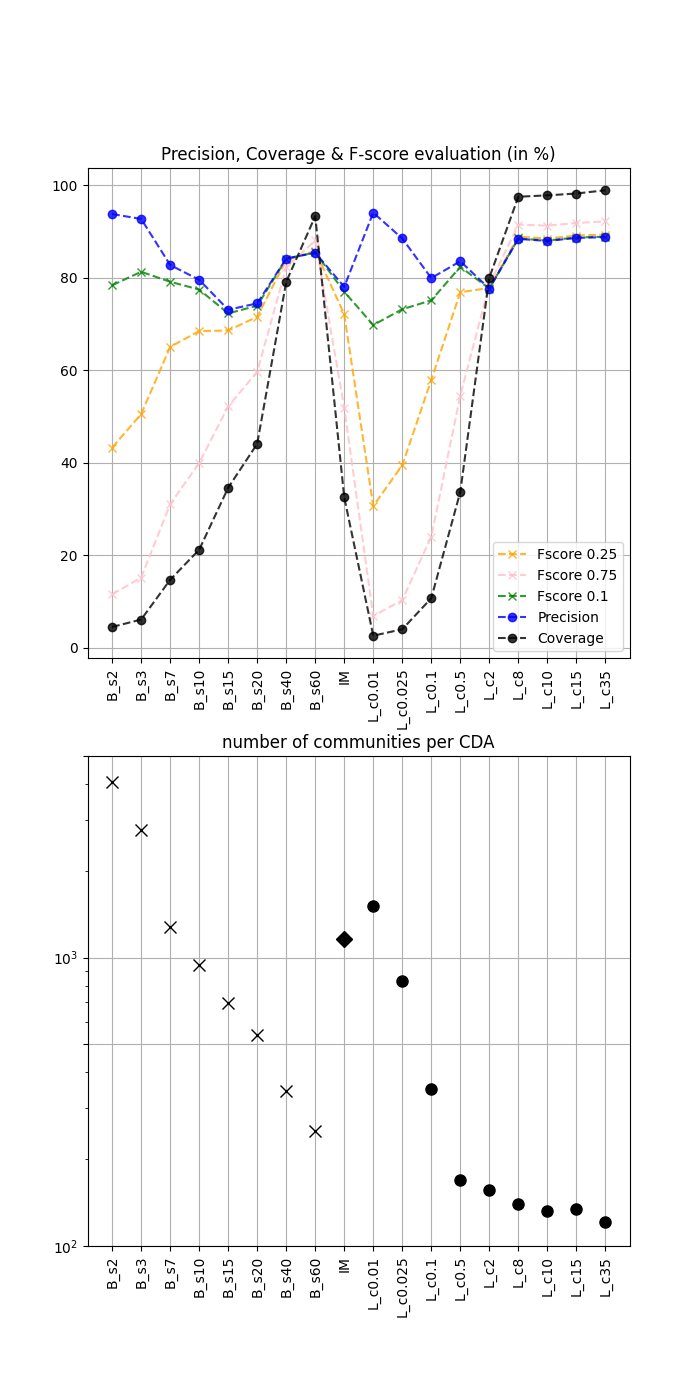}
    \caption{Lower panel: number of communities found by the CDA. Top panel: Precision (percentage of agreement between CDA and NLPCA categorization), Coverage (percentage of users covered by our 4 selected categories) and F-score (weighted score between precision and coverage)}
    \label{fig:nbcat}
\end{figure}

In Fig.\ref{fig:nbcat} (lower panel) we show the number of identified communities per CDA. Depending on the algorithm and on the parameters (if any), the number of communities can change by an order of magnitude. Hence the 4 selected (biggest) communities for each algorithm range from a few per cent of the total number of users, to most of them. This is what we refer to as the coverage (black curve) in the top panel of Fig.\ref{fig:nbcat}. This coverage can be compared with the \textit{precision} we previously considered (percentage of agreement between CDA and NLPCA in the classification of test set users). As one can expect, when the coverage is low, the precision is high because the narrative within small communities is not diverse. The precision decreases as the coverage increases, until a minimum is reached. Then we see the opposite trend for Louvain and BEC: precision and coverage increase together.

For instance, if one is interested in categorizing $80\%$ of users with a precision  of $90\%$ then we see that the best option is to use Louvain with parameter $c\ge 8$.

For group analysis in social science, the coverage of the users is most certainly of primary importance hence a good and natural score to use is the F-score,
\begin{equation}
 \rm F_\beta =(1+\beta^2) \frac{P\times R}{\beta^2 P +R}
\end{equation}

where $R$ is the recall, here corresponding to the coverage, $P$ the precision, and $\beta$ is a real factor chosen such that the recall is considered $\beta$ times as important as the precision . In Fig.\ref{fig:nbcat} (top panel) we display the F-score function weighting the coverage at $0.1$, $0.25$ and $0.75$. In such a case it is clear that the best performing CDA corresponds to the Louvain, with a best re-scaled modularity parameter $c\ge 8$, while the optimal scale s for BEC is $\ge 40$ (maximum of all F-scores).

The number of categories identified by the CDA is also something interesting to consider. In Fig.\ref{fig:nbcat} (lower panel) we see that the number of communities decreases when the coverage increases. Interestingly, there is a case where BEC, Infomap and Louvain have approximately the same number of communities: ($B_s^{7}; IM; L_c^{0.01}$). For this triplet, the Coverage is ($15\%; 33\%; 3\%$) while the Precision is ($83\%; 78\%;94\%$). This means that for Infomap we have a bigger clusters than for the Louvain, BEC being in between. 
Another interesting triplet is ($B_s^{7}$; $IM$; L$_c^{2}$) for which the F-score with weight 0.1 is similar and about $\sim 80\%$. Finally, for ($B_s^{15}$; $IM$; L$_c^{0.5}$) the coverage is the same but the precision ($82\%; 78\%;73\%$) shows that Louvain provides a better choice. 

\subsection{A Pseudo-Entropy measure of the NLPCA.}
\begin{figure}
    \centering
    \includegraphics[width=1.\linewidth]{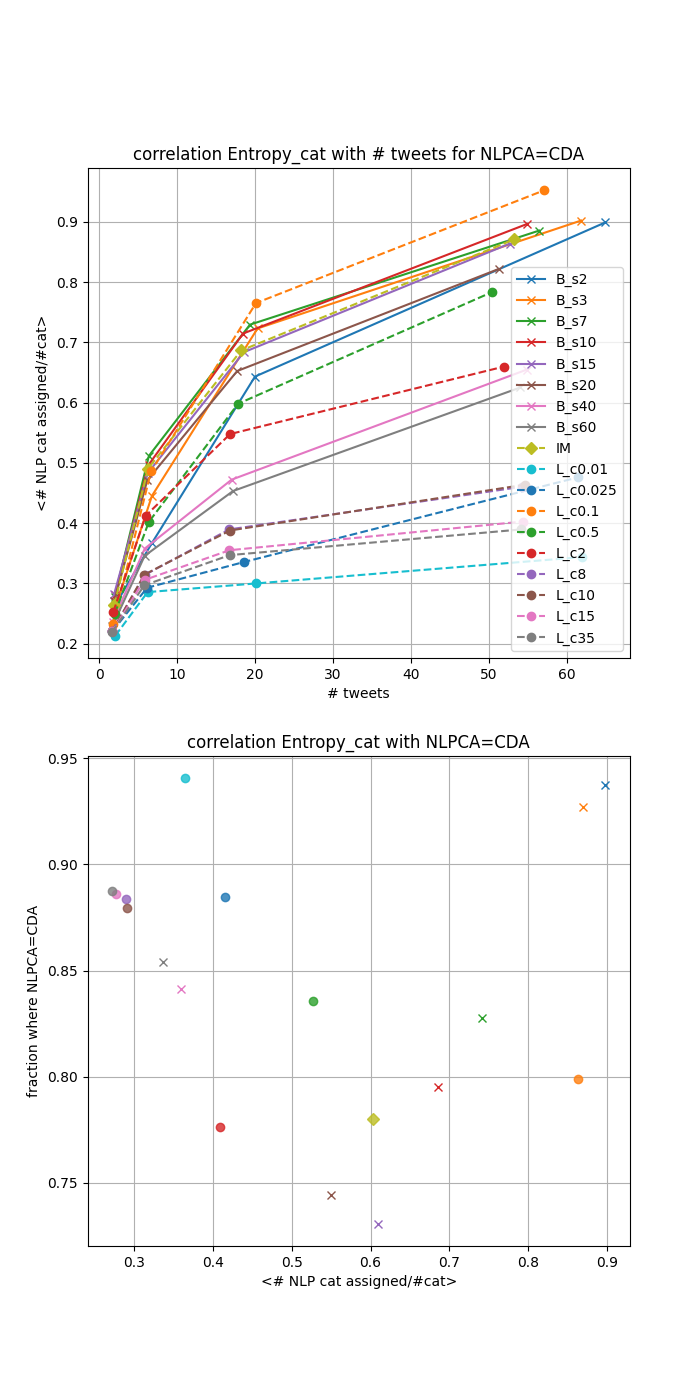}
    \caption{correlation between the entropy measure and the number of tweets (top panel) and the classification agreement with NLPCA (lower panel).}
    \label{fig:Fig2}
\end{figure}
Interestingly we may characterize the average number of distinct NLPCA categories for a user based on all his tweets. For instance if a user made 10 tweets, the NLPCA can assigns the 10 tweets to his CDA category, e.g. category 1, but it can also assign 5 tweets to category 1, 2 tweets to category 2 and 3, and 1 tweet to category 4. This provides a measure of the entropy of the categorization that we test for all CDA. If the entropy is null then it means that the NLP categorization of a user is without a doubt in the CDA category.

In Fig.~\ref{fig:Fig2} we display on the top panel the average of this entropy over all tested users as a function of their number of tweets. The more tweets a user has posted, the more likely it becomes for the NLPCA to assign a tweet to a different category than his CDA category. So for a fixed number of tweets we can compare the entropy of the different CDA. On the lower panel we display the fraction of users where the NLPCA categorization agrees with the CDA, as a function of this entropy measure.

Interestingly, the Louvain algorithm is the CDA algorithm that leads to the more stable NLPCA (users are assigned to a fewer number of categories) compared to the BEC CDA, while Infomap lies in the middle. This could be interpreted as a more subtle community structure in the BEC, where users are not necessarily central in their own community. This intuition is also confirmed by the coverage of users the CDA find. For BEC, the entropy is clearly related to the coverage of users: the entropy decreases as coverage increases. For the Louvain, this entropy is stable for $c\ge 8$, similar to what we find for the coverage in Fig.\ref{fig:nbcat}, while it increases when the coverage $\le 80\%$ which corresponds to $c<8$.  

Again, for the triplets case we previously considered with a similar number of communities ($B_s^{7}; IM; L_c^{0.01}$), ($B_s^{7}; IM; L_c^{2}$) where we have a similar F-score and ($B_s^{15}$; $IM$; L$_c^{0.5}$) where the coverage is the same, we observe that the entropy among these CDA is significantly different, suggesting different community properties for each CDA.

Finally, we have checked that the wrongly assigned users do not show distinguishable global properties on the network. For each CDA, we compare the distribution of the vector centrality of the wrongly assigned users to the distribution of all users without finding significant deviations. This is also true when comparing the CDA among them because the test set of users is mostly the same for all CDA. Among the two best performing CDA: $B_s^{60}$ and $L_c^{10}$ we find the $15\%$ and the $12\%$ of wrongly assigned users in both CDA (respectively), are about half the same users.
This category of users are either "indecisive" or easily influenced, and can be an interesting social group to study in their own right.

\section{Comparison analysis of the CDA in the light of NLP\label{sec_indiv}}
\begin{figure}
        \centering
        \includegraphics[width=1\linewidth]{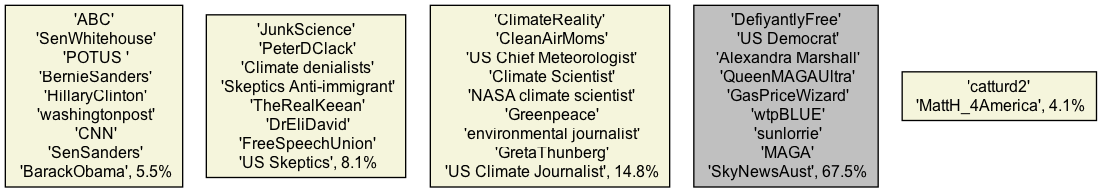}
        \caption{Infomap community structures for the selected users. Yellow boxes correspond to one of our four categories while the grey box correspond to the catch-all-term category. The percentage in each box corresponds to the fraction of all users in each of the category. }
         \label{fig:denI}
    \end{figure}
   
\begin{figure}
    \centering
    \includegraphics[width=1\linewidth]{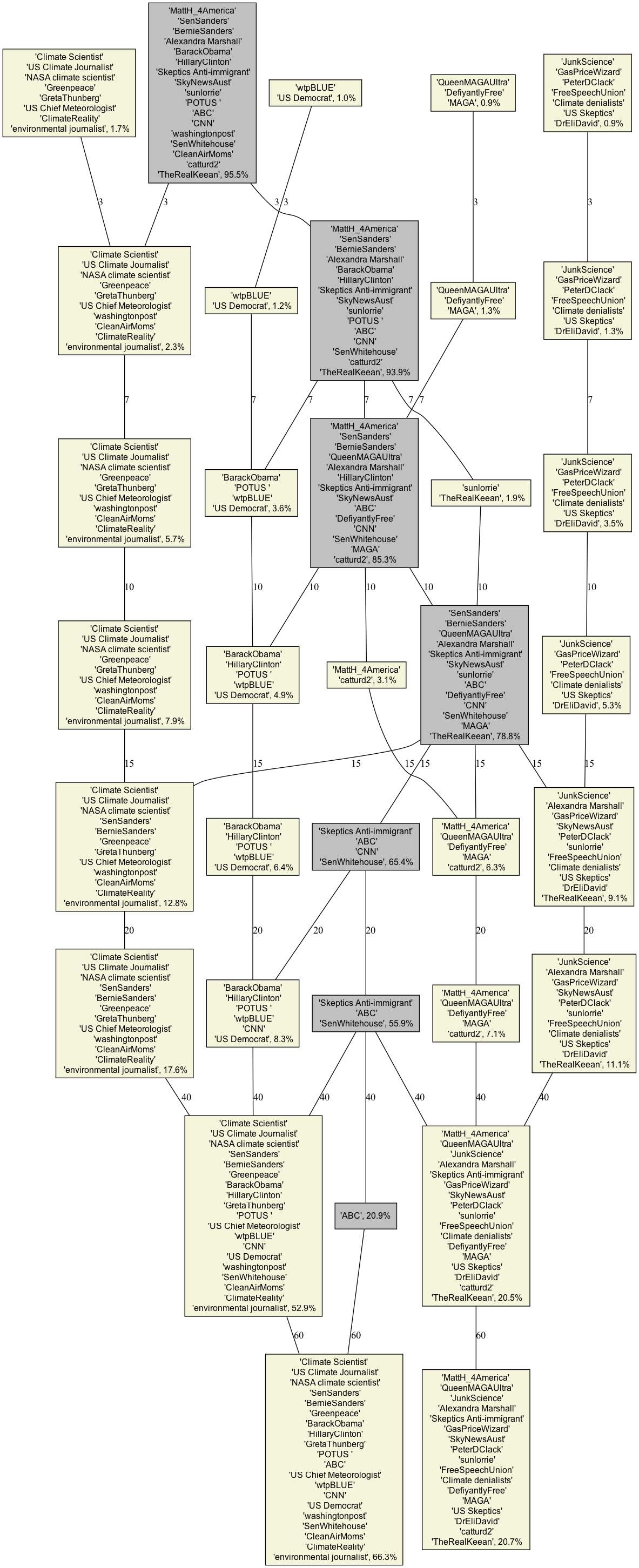}
    \caption{BEC Dendogram of our selected users community as function of the BEC parameters s (displayed on the edges). Yellow boxes correspond to one of our 4 categories while the grey box correspond to the catch-all-term category. The percentage in each box corresponds to the fraction of all users in each of the category.} \label{fig:denB}
    \end{figure}
    
\begin{figure}
    \centering
    \includegraphics[width=0.75\linewidth]{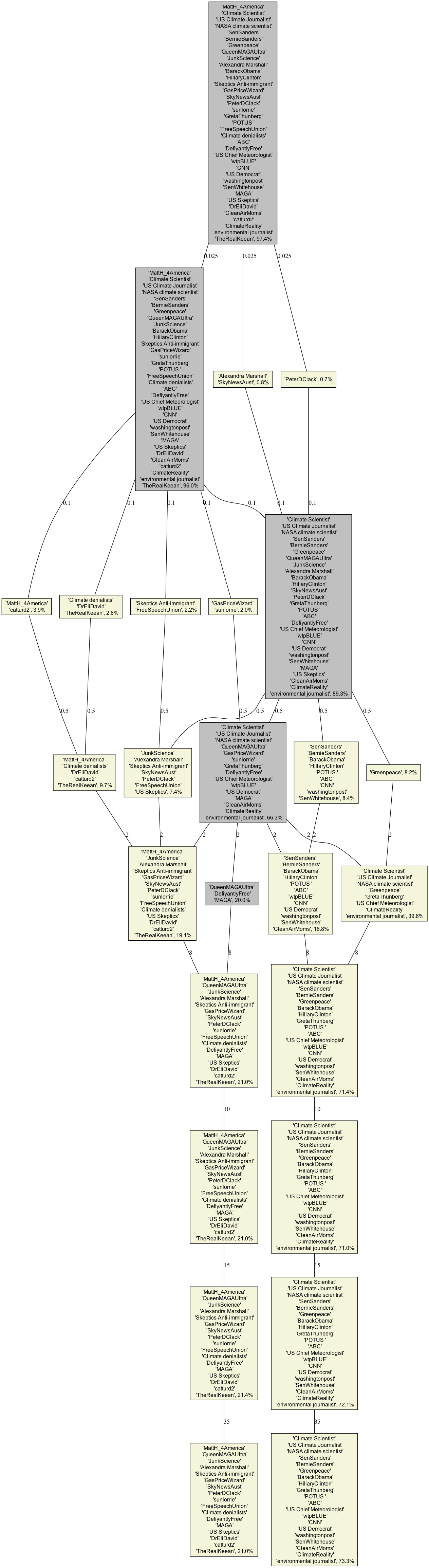}
    \caption{Louvain Dendogram of our selected users community as function of the rescaled modularity parameter c (displayed on the edges). Yellow boxes correspond to one of our 4 categories while the grey box correspond to the catch-all-term category. The percentage in each box corresponds to the fraction of all users in each of the category.}
    \label{fig:denL}
\end{figure}

We now turn to a more refined analysis of the CDA categories by flagging influential users that belong to them. Those are the users from the training set that are public figures, or represent political parties, media, or are influencers. We also report a few users that have tens to hundreds of thousands of followers with a strong view on climate change. Among the pro-climate users we have $CleanAirMoms$ 37,000 followers, a community of `moms and dads who are uniting for clean air and our kids’ health'; $wtpBLUE$, 12,000 followers, and a `grassroots GOTV organization dedicated to electing Democrats'. Among the American denialists we select $MattH_4America$, 100,000 followers defined as `America First - Patriot '; $Catturd$, 2.5M followers, a MAGA influencer; $DrEliDavid$, 620,000 followers, entrepreneurs; for the Australian denialists: $Alexandra Marshall$ 77,300 followers, writer/artist; $PeterDClarck$, 37,000 followers, journalist advocating for carbon emission. In the Canadian denialists we have: $GasPriceWizard$, 54,000 followers, a former liberal MP; $Sunlorrie$, 108,000 followers, journalist; $TheRealKeean$, 300,000 followers, a journalist.

\subsection{Community structures for the same coverage.}

Infomap does not provide a scale parameter, the four biggest communities cover $32.5\%$ of all users. These communities,as displayed in Fig.\ref{fig:denI}, are easy to interpret. The largest ($14.8\%$) corresponds to pro-climate activists, scientist and media. The second ($8.1\%$) corresponds to climate denialists that focus on denying that CO$_2$ and fossil fuel are the cause of climate change. The third ($5.5\%$) corresponds to Democrats and some mainstream media, while the fourth ($4.1\%$) corresponds to communities around MAGA influencers for whom climate change is a``hoax" or a``cult". 

We can compare these communities to what Louvain and BEC provide for about the same coverage (for $c=0.5$ and $s=15$, respectively). Interestingly, they differ in the way they categorize the top influencers. On the denialist side, Infomap and BEC tend to agree on the cluster of clear MAGA supporter and on another more dedicated to``experts" and lobbies, while Louvain places some big influencer from the lobbies cluster into the MAGA cluster. 

The situation is different for the pro-climate communities. Here, Infomap and Louvain agree to make a environmental NGO/activits cluster and another with Democrat leaders such as Biden, Obama, Hilary Clinton and Bernie Sanders, while BEC includes Bernie Sanders in the environmental NGO/activits cluster. Both clustering make sense but we might argue that from the point of view of the climate debate, Bernie Sanders was indeed closer to  NGO/activits than the mainstream Democrats in his public statements, which is reflected in the BEC clustering. 

From a narrative perspective, the precision provides a measure of how homogeneous the textual content of these communities is. For the same coverage, the Louvain provides the best precision ($\sim 82\%$), followed by Infomap ($\sim 78\%$) and BEC ($\sim 73\%$). 

\subsection{Community structure evolution for different parameters.}
The analysis of the evolution of community structures as a function of the scale parameters $s$ for BEC (Fig.~\ref{fig:denB}) and $c$ for Louvain (Fig.~\ref{fig:denL}) provides more insight into these differences. Similarly to the coverage evolution, for small values of $s$ or $c$ we have many small communities, and hence most users belong to the catch-all-term category (displayed in grey). As the values of these parameters increase, the number of communities decreases and the fraction of users in our four categories increases. For our selection of users we observed that they merge into only two groups at high scales: $s\ge 60$ and $c \ge 8$, for BEC and Louvain respectively (the other two categories do not contain our selected influencers). These two groups are the same for the two CDA and can be interpreted as climate denialists and pro-climates. In these cases, the coverage $>85\%$ and the precision $>85\%$ for the two CDA becomes similar.

Interestingly, BEC identifies well the four categories of opinion among the top four communities even for low scale resolution (community sizes $<1\%$) and integrates more actors as the scale is increased, Louvain focuses on peripheral communities and important ones appear only for medium scales (community sizes $2-15\%$). This is not surprising because optimizing modularity leads to merging small communities into larger ones, even when those small communities are well defined and weakly connected to one another \cite{Kumpula_2007}. Thus, we only see the four categories (given our selected users), when $c=0.1$. Moreover, it seems that BEC faithfully reflects the structure of positions on climate, with an initial integration of Bernie Sanders' current into the community of pro-climate activists, followed by a merger of all pro-climates into a single community for high values of $s$.

\section{Conclusion \& Discussion}\label{sec_conclu}

The study of social networks has experienced significant growth, leading to substantial advancements in understanding the dynamics of social structures and interactions \cite{borgatti_analyzing_2011},\cite{lazer_computational_2009}, \cite{newman_structure_2003}. When analysing social networks, CDA are key tools that allow to reduce the analysis of a complex network system with exchanges among many users to a bigger picture where we study the exchanges between communities: we quantify the structure of large complex networks in terms of a smaller number of cohesive components. 

These components are densely connected but does not make the assumptions that users within produce a similar cultural production. Indeed, CDA require edges/links to define the relationship between the nodes (users) rather than their cultural production. We assume in this work that users within a given community share a similar opinion. To run a CDA on a network, the edges/links (relationship) between users can be dynamic (such as retweets) or static (such as a list of followers). By considering the entangled nature of the produced social network communities and the cultural production of the users, we derive a new metric to asses the quality of the CDA. This metric is not based upon the network properties but on the coherence of the narratives within a community. It allows us to identify optimal parameters of the CDA and make a direct comparison for the precision of different CDA. 

In addition we find an interesting application in classifying users via their textual publication. Once we have an optimal CDA we can further predict with high accuracy ($\sim >85\%$) the opinion of a new user with only a few sentences, by performing a supervised training with NLPCA.

Finally, this work shows interesting features of `wrongly assigned communities': by taking the intersection of users that were assigned to a different category between a couple (or more) CDA and NLPCA, we can identify these users as being either `indecisive' or easily influenced, as they produce textual information that is inconsistent with their CDA communities. These users are an interesting social group that can be studied in the framework of opinion dynamics, as they are likely bridges between distinct communities.

\begin{acks}
This work was supported by the Complex Systems Institute of Paris Île-de-France (ISC-PIF) and the EU NODES project (LC-01967516).
\end{acks}

\bibliographystyle{SageV}
\bibliography{biblio}

\begin{thebibliography}{10}
\providecommand{\url}[1]{\texttt{#1}}
\providecommand{\urlprefix}{URL }
\expandafter\ifx\csname urlstyle\endcsname\relax
  \providecommand{\doi}[1]{DOI:\discretionary{}{}{}#1}\else
  \providecommand{\doi}{DOI:\discretionary{}{}{}\begingroup \urlstyle{rm}\Url}\fi
\providecommand{\eprint}[2][]{\url{#2}}

\bibitem{golbeck2011predicting}
Golbeck J, Robles C, Edmondson M et~al.
\newblock Predicting personality from twitter.
\newblock In \emph{2011 IEEE Third International Conference on Privacy, Security, Risk and Trust and 2011 IEEE Third International Conference on Social Computing}. IEEE, pp. 149--156.

\bibitem{gosling2011manifestations}
Gosling SD, Augustine AA, Vazire S et~al.
\newblock Manifestations of personality in online social networks: Self-reported facebook-related behaviors and observable profile information.
\newblock \emph{Cyberpsychology, Behavior, and Social Networking} 2011; 14(9): 483--488.

\bibitem{GAREY1976237}
Garey M, Johnson D and Stockmeyer L.
\newblock Some simplified np-complete graph problems.
\newblock \emph{Theoretical Computer Science} 1976; 1(3): 237--267.
\newblock \doi{https://doi.org/10.1016/0304-3975(76)90059-1}.
\newblock \urlprefix\url{https://www.sciencedirect.com/science/article/pii/0304397576900591}.

\bibitem{chakraborty2016metrics}
Chakraborty T, Dalmia A, Mukherjee A et~al.
\newblock Metrics for community analysis: A survey, 2016.
\newblock \eprint{1604.03512}.

\bibitem{gaume2024unified}
Gaume B, Achitouv I and Chavalarias D.
\newblock A unified graph clustering framework for complex systems modeling.
\newblock \emph{SSRN} 2024; \doi{10.2139/ssrn.4766265}.
\newblock \url{http://dx.doi.org/10.2139/ssrn.4766265}.

\bibitem{yang2012defining}
Yang J and Leskovec J.
\newblock Defining and evaluating network communities based on ground-truth, 2012.
\newblock \eprint{1205.6233}.

\bibitem{Radicchi_2004}
Radicchi F, Castellano C, Cecconi F et~al.
\newblock Defining and identifying communities in networks.
\newblock \emph{Proceedings of the National Academy of Sciences} 2004; 101(9): 2658–2663.
\newblock \doi{10.1073/pnas.0400054101}.
\newblock \urlprefix\url{http://dx.doi.org/10.1073/pnas.0400054101}.

\bibitem{Jianbo}
Shi J and Malik J.
\newblock Normalized cuts and image segmentation.
\newblock \emph{IEEE Transactions on Pattern Analysis and Machine Intelligence} 2000; 22(8): 888--905.
\newblock \doi{10.1109/34.868688}.

\bibitem{NewmanGirvan}
Newman MEJ and Girvan M.
\newblock Finding and evaluating community structure in networks.
\newblock \emph{Physical Review E} 2004; 69(2).
\newblock \doi{10.1103/physreve.69.026113}.
\newblock \urlprefix\url{http://dx.doi.org/10.1103/PhysRevE.69.026113}.

\bibitem{Lancichinetti_2008}
Lancichinetti A, Fortunato S and Radicchi F.
\newblock Benchmark graphs for testing community detection algorithms.
\newblock \emph{Physical Review E} 2008; 78(4).
\newblock \doi{10.1103/physreve.78.046110}.
\newblock \urlprefix\url{http://dx.doi.org/10.1103/PhysRevE.78.046110}.

\bibitem{frank2018inferring}
Frank MR, Sun L, Cebrian M et~al.
\newblock Inferring user demographics from social media text.
\newblock \emph{Proceedings of the National Academy of Sciences} 2018; 115(25): 7697--7702.
\newblock \urlprefix\url{https://www.pnas.org/content/115/25/7697}.

\bibitem{van2018gender}
Van~Hee C, Jacobs G, Wijnhoven W et~al.
\newblock Gender classification and bias mitigation in twitter analytics.
\newblock \emph{Social Network Analysis and Mining} 2018; 8(1): 50.
\newblock \urlprefix\url{https://link.springer.com/article/10.1007/s13278-018-0501-3}.

\bibitem{schwartz2013personality}
Schwartz HA, Eichstaedt JC, Kern ML et~al.
\newblock Personality, gender, and age in the language of social media: The open-vocabulary approach.
\newblock \emph{PLOS ONE} 2013; 8(9): e73791.
\newblock \urlprefix\url{https://journals.plos.org/plosone/article?id=10.1371/journal.pone.0073791}.

\bibitem{ferrara2014user}
Ferrara MA, De~Meo P, Ferrara E et~al.
\newblock User classification in online social networks.
\newblock \emph{ACM Computing Surveys (CSUR)} 2014; 47(3): 21.
\newblock \urlprefix\url{https://dl.acm.org/doi/10.1145/2700489}.

\bibitem{le2019detecting}
Le~Mens G and Vedres B.
\newblock Detecting influencers in online social networks: The role of individual communicative factors.
\newblock \emph{Social Networks} 2019; 56: 15--28.
\newblock \urlprefix\url{https://www.sciencedirect.com/science/article/pii/S0378873318301670}.

\bibitem{chavalarias_new_2023}
Chavalarias D, Bouchaud P, Chomel V et~al.
\newblock The new fronts of denialism and climate skepticism, 2023.
\newblock \urlprefix\url{https://hal.science/hal-04103183}.

\bibitem{Blondel_2008}
Blondel VD, Guillaume JL, Lambiotte R et~al.
\newblock Fast unfolding of communities in large networks.
\newblock \emph{Journal of Statistical Mechanics: Theory and Experiment} 2008; 2008(10): P10008.
\newblock \doi{10.1088/1742-5468/2008/10/p10008}.
\newblock \urlprefix\url{http://dx.doi.org/10.1088/1742-5468/2008/10/P10008}.

\bibitem{Lambiotte_2014}
Lambiotte R, Delvenne JC and Barahona M.
\newblock Random walks, markov processes and the multiscale modular organization of complex networks.
\newblock \emph{IEEE Transactions on Network Science and Engineering} 2014; 1(2): 76–90.
\newblock \doi{10.1109/tnse.2015.2391998}.
\newblock \urlprefix\url{http://dx.doi.org/10.1109/TNSE.2015.2391998}.

\bibitem{Rosvall_2008}
Rosvall M and Bergstrom CT.
\newblock Maps of random walks on complex networks reveal community structure.
\newblock \emph{Proceedings of the National Academy of Sciences} 2008; 105(4): 1118–1123.
\newblock \doi{10.1073/pnas.0706851105}.
\newblock \urlprefix\url{http://dx.doi.org/10.1073/pnas.0706851105}.

\bibitem{achitouv2023natural}
Achitouv I, Gorduza D and Jacquier A.
\newblock Natural language processing for financial regulation, 2023.
\newblock \eprint{2311.08533}.

\bibitem{Vaswani}
Vaswani A, Shazeer N, Parmar N et~al.
\newblock Attention is all you need.
\newblock \emph{Advances in NeurIPS} 2017; 30.

\bibitem{BERT}
Devlin J, Chang MW, Lee K et~al.
\newblock Bert: Pre-training of deep bidirectional transformers for language understanding, 2018.
\newblock \url{https://arxiv.org/abs/1810.04805}.

\bibitem{nguyen2020bertweet}
Nguyen DQ, Vu T and Nguyen AT.
\newblock Bertweet: A pre-trained language model for english tweets, 2020.
\newblock \eprint{2005.10200}.

\bibitem{bottou2010large}
Bottou L.
\newblock Large-scale machine learning with stochastic gradient descent.
\newblock In \emph{Proceedings of COMPSTAT'2010}. Physica-Verlag HD, pp. 177--186.

\bibitem{cortes1995support}
Cortes C and Vapnik V.
\newblock Support-vector networks.
\newblock \emph{Machine Learning} 1995; 20(3): 273--297.

\bibitem{rumelhart1986learning}
Rumelhart DE, Hinton GE and Williams RJ.
\newblock Learning representations by back-propagating errors.
\newblock \emph{Nature} 1986; 323(6088): 533--536.

\bibitem{breiman2001random}
Breiman L.
\newblock Random forests.
\newblock \emph{Machine Learning} 2001; 45(1): 5--32.

\bibitem{shaw1954group}
Shaw ME.
\newblock Group structure and the behavior of individuals in small groups.
\newblock \emph{The Journal of psychology} 1954; 38(1): 139--149.

\bibitem{Kumpula_2007}
Kumpula JM, Saramäki J, Kaski K et~al.
\newblock Limited resolution in complex network community detection with potts model approach.
\newblock \emph{The European Physical Journal B} 2007; 56(1): 41–45.
\newblock \doi{10.1140/epjb/e2007-00088-4}.
\newblock \urlprefix\url{http://dx.doi.org/10.1140/epjb/e2007-00088-4}.

\bibitem{borgatti_analyzing_2011}
Borgatti SP and Halgin DS.
\newblock \emph{Analyzing Social Networks}.
\newblock SAGE Publications, 2011.

\bibitem{lazer_computational_2009}
Lazer D, Pentland AS, Adamic L et~al.
\newblock Computational social science.
\newblock \emph{Science} 2009; 323(5915): 721--723.

\bibitem{newman_structure_2003}
Newman ME.
\newblock The structure and function of complex networks.
\newblock \emph{SIAM Review} 2003; 45(2): 167--256.

\end{thebibliography}

\end{document}